 \documentclass[twocolumn]{emulateapj}

\shorttitle {Three M31 GCs, including a new BHC}
\shortauthors{Barnard et al.}

\begin{document}


\title{New XMM-Newton analysis of three bright X-ray sources in M31 globular clusters, including a new black hole candidate}


\author{R. Barnard\altaffilmark{1}, H. Stiele\altaffilmark{2}, D. Hatzidimitriou\altaffilmark{3,4}, A.K.H. Kong\altaffilmark{5}, B.F. Williams\altaffilmark{6}, W. Pietsch\altaffilmark{2}, U.C. Kolb\altaffilmark{1},\\  F. Haberl\altaffilmark{2}, G. Sala\altaffilmark{2}}





\altaffiltext{1}{The Open University, Walton Hall, Milton Keynes UK, MK7 6AA. email: r.barnard@open.ac.uk}
\altaffiltext{2}{Max-Planck-Institut f\"ur Extraterrestrische Physik,\\ Garching, Germany}
\altaffiltext{3}{University of Crete}
\altaffiltext{4}{Foundation for Research and Technology, Hellas}
\altaffiltext{5}{Institute of Astronomy and Department
of Physics,\\ National Tsing Hua University, Hsinchu, Taiwan. }
\altaffiltext{6}{University of Washington Astronomy Department, Box 351580, Seattle, WA}


\begin{abstract}
We present detailed analysis of three globular cluster X-ray sources in the XMM-Newton extended survey of M31. 
The X-ray counterpart to the M31 globular cluster Bo\thinspace 45 (XBo\thinspace 45) was observed with XMM-Newton on 2006 December 26. Its combined pn+MOS 0.3--10 keV lightcurve was seen to vary by $\sim$10\%, and its 0.3--7.0 keV emission spectrum was well described by an absorbed power law with photon index 1.44$\pm$0.12. Its variability and emission is characteristic of low mass X-ray binaries (LMXBs) in the low-hard state, whether the accretor is a neutron star or black hole. Such behaviour is typically observed at luminosities $\la$10\% Eddington. However, XBo\thinspace 45 exhibited this behaviour at an unabsorbed, 0.3--10 keV luminosity of { 2.5$\pm$0.2$\times 10^{38}$} erg s$^{-1}$, or{ $\sim$140\%} Eddington for a 1.4 $M_{\odot}$ neutron star accreting hydrogen. Hence, we identify  XBo\thinspace 45 as a  new candidate black hole LMXB. { XBo\thinspace 45 appears to have been consistently bright for $\sim$30 years, consistent with theoretical prediction for a globular cluster black hole binary formed via tidal capture}.
 Bo\thinspace 375 was observed in the 2007, January 2 XMM-Newton observation, and has a two-component spectrum that is typical for a bright neutron star LMXB. 
 Bo\thinspace 135 was observed in the same field as Bo\thinspace 45, and could contain either a black hole or neutron star. 
\end{abstract}


\keywords{X-rays: binaries --- galaxies: individual: M31 --- methods: data analysis}



\section{Introduction}

Of the 14 bright low mass X-ray binaries (LMXBs) associated with Galactic globular clusters (GCs), 13 are known to contain neutron stars, and the other one probably does as well \citep[see e.g.][ and references within]{wa01,int04}.
 However, black hole binaries have been seen in extragalactic GCs, proving that they form in such environments.   For
example, \citet{mac07} discovered the first  GC black hole binary in the giant elliptical galaxy NGC 4472. 

{ The lack of known GC black hole binaries has been a subject of great interest for theoretical modellers.
 \citet{kal04} show that black hole binaries formed through exchange interactions should have duty cycles of $\sim$0.001, consistent with the absence of GC black hole binaries at the  time. They also predict that black hole binaries formed  through tidal capture of a main sequence star would be bright, persistent X-ray sources, and infer  from their absence that tidal capture probably disrupts the main sequence star. However, neither of these results preclude the discovery of new GC black hole binaries.

}

The bright X-ray sources of M31 have been studied for over 25 years, with Einstein \citep{tf91}, ROSAT \citep{S97,S01}, Chandra \citep[see e.g.][]{dis02, kaa02,will04} and XMM-Newton \citep[see e.g.][]{shi01,trud05,pfh05,lsg08}.

\citet{dis02} conducted a Chandra survey of selected regions of  M31, and found that most of their bright X-ray sources were associated with GCs, with $\sim$10\% of GC sources exhibiting 0.5--7 keV luminosities $\ga$10$^{38}$ erg s$^{-1}$. They also showed the M31 GC  population to be significantly different from that of the Milky Way, as $\sim$30\% of the M31 GC X-ray sources exhibited luminosities $>$10$^{37}$ erg s$^{-1}$, whereas  only one out of 11  Galactic GC X-ray sources exceeded 10$^{37}$ erg s$^{-1}$ \citep{verb95}. These results contradict those of \citet{S01}, who concluded that the GC X-ray populations of the Milky Way and M31 were similar.

\citet{dis02} also compared the optical properties of M31 GCs with and without X-ray sources, as well as M31 GCs and Milky Way GCs with X-ray sources. They compared optical colours and apparent magnitudes, radial velocities, metallicities and colour excesses, as well as core radii. They found the only possibly significant difference to be in their luminosities: the median luminosity of M31 GCs with X-ray sources was $\sim$0.55 magnitudes higher than that of M31  GCs without. Similarly, M31 X-ray GCs were slightly brighter in V than Milky Way X-ray GCs. \citet{dis02} { speculated} that this might associate X-ray sources with  higher GC masses. 

 We report on three bright X-ray sources in M31, all associated with populous, old globular clusters: Bo\thinspace 45, Bo\thinspace 135 and Bo\thinspace 375. 
We will argue that Bo\thinspace 45 contains a good black hole (BH) candidate, while Bo\thinspace 375 is likely to contain a neutron star (NS).  Bo\thinspace 135 is more likely to contain a NS, but we cannot rule out a BH accretor.

We first review the emission and variability of LMXBs in their various states in Section~\ref{bhs}, and then discuss  known properties of our  targets and their host clusters in Section~\ref{gc}.
 We describe the observation and analysis in Section~\ref{obs}, and our results in Section~\ref{res}. We finally present our discussion and conclusions in Section~\ref{dis}.

\section{Variability and emission spectra of  LMXBs in different states}
\label{bhs}

It was shown by \citet{vdk94} that NS and BH  LMXBs accreting at low rates exhibit remarkably similar emission spectra and variability. In this low-hard state, the power density spectrum (PDS) may be characterised by a broken power law with spectral index $\gamma$ changing from $\sim$0 to $\sim$1 at some break frequency; also, the r.m.s. variability is high \citep[$\sim$10--50\%, ][]{vdk94, vdk95,mr03}. However, at higher accretion rates, the PDS may be described by a simple power law  with $\gamma$ $\sim$1 and the r.m.s.  variability is  only a few percent \citep[][]{vdk94, vdk95,mr03}.

This change in variability from low to high accretion rate is accompanied by a change in the emission spectra of LMXBs: low-state LMXBs have hard spectra that are characterised by power laws with photon index $\sim$1.4--1.7, regardless of the primary \citep{vdk94,mr03}. Higher accretion rates lead to distinctive emission spectra, depending on the accretor. BH LMXBs exhibit a thermally dominated state (also known as the high/soft state) where a 0.7--2 keV disc blackbody contributes $\ga$90\% of the 0.3--10 keV flux \citep{mr03}; they also exhibit a ``steep power law'' state, where an additional power law component is observed, with a photon index $>2.4$ \citep{mr03}. Meanwhile,  the emission spectra of high accretion rate NS LMXBs remain predominantly non-thermal, although a second, blackbody component becomes increasingly important at higher luminosities, contributing up to 50\% of the flux \citep[see e.g.][and references within]{wsp88,cbc95,cbc01,bcb03}. 

Van der Klis (1994) proposed that  LMXBs switched from low to high accretion rate behaviour at some constant fraction of the Eddington limit. \citet{bok03} realised that if this were true, then BH LMXBs would be capable of exhibiting low-state variability and spectra at higher luminosities than neutron star LMXBs, as the Eddington limit is proportional to the mass of the accretor. { \citet{bko04} found some empirical evidence for a transition at $\sim$10\% Edddington}.  We would therefore expect a LMXB containing a $\sim$10 M$_{\odot}$ BH to exhibit low-state characteristics at $\sim$10 times the highest luminosity of low-states observed in 1.4 M$_{\odot}$ NS LMXBs. 

{
\citet{glad07} have systematically analysed  observations of 40 disc-accreting neutron stars (atoll class LMXBs and millisecond pulsars) from the RXTE database, in order to follow their spectral evolution as a function of luminosity. They excluded  sources with high line-of-sight or intrinsic absorption (equivalent to $>$3$\times$10$^{22}$ H atom cm$^{-2}$), as well as those sources which readily exceed the Eddington limit such as Sco X-1 \citep[see e.g.][]{bcb03}. They found two different types of hard/soft transition: those where the
spectrum softens at all energies, leading to a diagonal track on a
colour-colour diagram, and those where only the higher energy spectrum
softens, giving a vertical track. The  diagonal transitions were made at $\la$10\% Eddington, while  vertical transitions occurred at $\la$2\% Eddington in the 0.01--1000 keV band.

Furthermore, GS\thinspace 2023+338 (a.k.a. V404 Cygni) contains a BH primary with mass $\sim$12 M$_{\odot}$ \citep{shab94}; it has exhibited low-state PDS and emission spectra at 2--37 keV luminosities $>$3$\times$10$^{38}$ erg s$^{-1}$  \citep{miy92,oost97}. This corresponds to a luminosity  of  $>$1.2$\pm$0.4$\times$10$^{38}$ erg s$^{-1}$  in the XMM-Newton pass band (0.3--10 keV), for power law emission with photon index 1.0--1.4, as found by \citet{miy92}. This translates as  8\% Eddington for a 12 M$_{\odot}$ black hole accretor, suggesting that spectral transitions in disc-accreting black holes may occur at similar fractions of the Eddington limit.
}

We note that the M31 BH candidates identified by \citet{bok03,bko04} were contaminated by artificial variability that was introduced when combining non-synchronised lightcurves from the different instruments on board XMM-Newton \citep{bs07}. The current work employs corrected techniques { described in \citet{bs07}}.

\section{Known properties of Bo\thinspace 45,  Bo\thinspace 135 and Bo\thinspace 375}
\label{gc}

\subsection{X-ray properties}

{ The X-ray} counterpart to Bo\thinspace 45,  has not been previously observed by XMM-Newton or Chandra. However, it was observed by   Einstein in 1979--1980 \citep{tf91} and by ROSAT in the summers of 1991 and 1992 \citep{S97,S01}. \citet{S97} designated the X-ray source  as RX J0041.7+4134, and found  its 0.2--4.0 keV flux to have increased by a factor of $\sim$2 with respect to the Einstein observation, from 8.7$\pm$1.1 to 14.4$\pm$0.5 $\times 10^{-13}$ erg cm$^{-2}$ s$^{-1}$; this equates to a change in 0.2--4.0 keV luminosity from $\sim$6$\times$10$^{37}$  to $\sim$1.0$\times$10$^{38}$ erg s$^{-1}$ for a {  $\sim$780 kpc distance \citep{sg98}}. For this estimate, they assumed  a 5 keV bremsstrahlung emission model, suffering absorption equivalent to 7$\times 10^{20}$ H atom cm$^{-2}$, following \citet{tf91}. However, they reported no variability between ROSAT observations greater than 3$\sigma$.  { We shall refer to this source as XBo\thinspace 45 for the remainder of the paper}.

{ The X-ray source}  associated with Bo\thinspace 135  has been observed by both XMM-Newton and Chandra before; \citet{tp04} have analysed these data as part of a spectral survey of M31 GCs.  They found the 0.3--10 keV luminosity of  Bo\thinspace 135 to vary over 3.3--4.1$\times$10$^{38}$ erg s$^{-1}$, for power law spectra with photon indices varying over 1.48--1.66. { We name this source XBo\thinspace 135}.

 {  The X-ray source associated with Bo 375} has not been previously observed with XMM-Newton. However, it has been observed several times since 1979, with Einstein, ROSAT, ASCA and Chandra; \citet{dis02} have analysed many of these observations. They obtained good fits with absorbed power law models for the ASCA and ROSAT spectra, but not the Chandra ACIS-S spectrum; they obtained 0.5--2.4 keV luminosities of $\sim$1.4--2.3$\times$10$^{38}$ erg s$^{-1}$, assuming a distance of 780 kpc.  \citet{dis02}  modeled the ACIS-S spectrum with a 0.80 keV blackbody and a power law with photon index, $\Gamma$ = 1.67, for a 0.3--7 keV luminosity of 4.2$\times$10$^{38}$ erg s$^{-1}$. \citet{tp04} re-analysed this Chandra observation, as well as an additional ACIS-S observation taken $\sim$4 months later. They modeled the 0.3--7 keV spectra with absorbed power laws, and found that the later observation was a factor $\sim$2 brighter.  { We name this source XBo 375}.
 
\subsection{Optical properties of the host clusters}

 Bo\thinspace 45,  Bo\thinspace 135, and Bo\thinspace 375, the host clusters,  have been
 classified as {\em certain} globular clusters in the "The Revised
 Bologna Catalogue (RBC) of M31 globular clusters and candidates". They
 have been imaged with the HST and their nature has been firmly
 confirmed \citep[RBC V.3.4, January 2008;][]{gall04,gal05,gal06,gal07}.

 The integrated magnitudes and colours of the clusters can be
 found in RBC V.3.4 and can be used to provide estimates of the ages
 and masses of the clusters, in conjunction with reddening and
 distance and metal abundance values. Bo\thinspace 45 has $V$ = 15.78, $V-I$ = 1.27, while Bo\thinspace 135 has
 $V$ = 16.04, $V-I$ = 1.22, and Bo\thinspace 375 has $V$ = 17.61, $V-I$ = 1.02.
{  We have adopted a
 distance modulus of $(m-M)_o=24.471\pm 0.035 \pm  0.045$ \citep[784$\pm$13$\pm$17 kpc, ][]{sg98}; the first uncertainty is statistical and the second uncertainty is systemmatic}. We use a  reddening of
 $E(B-V)=0.10\pm0.03$, which is the average of the reddenings of all
 M31 clusters in \citet{ric05}. Actually, for Bo\thinspace 45, there are
 specific reddening measurements that have been tabulated in \citet{ric05}, yielding an average of  { $E(B-V)=0.10\pm0.06$}, which is identical
 to  the mean value we have adopted. Using these values, we have
 derived $M_V=-9.0, -8.74, -7.17$ and $(V-I)_o=1.13, 1.08, 0.88$ for Bo\thinspace 45, Bo
 135 and Bo\thinspace 375, respectively.{ The metal abundances of the three clusters are estimated to be $Z$=0.002, 0.0004 and 0.0012 respectively \citep{bel95}. According to \citet{sar07}, the above 
 values for $(V-I)_o$, $M_V$ and $Z$ suggest that Bo\thinspace 45 and Bo\thinspace 135 are at least as old as $\sim$10$^{10}$ yr and
 at least as massive as  $\simeq 10^6$ M$_{\odot}$, while the fainter optical luminosity and bluer color of Bo\thinspace 375 allow it
 to be as young as 1.2 Gyr and to have a mass as low as 7$\times$10$^4$
 M$_{\odot}$.}

 In conclusion,  both Bo\thinspace 45 and Bo\thinspace 135 seem to belong to the old massive
 globular cluster population of M31. This is not an unexpected
 result, since low mass X-ray binaries are  found preferentially in
 luminous (massive) and red globular clusters \citep{kun08}. However, Bo\thinspace 375 appears to be considerably less massive, and younger, than the other two GCs.

\section{Observations and data analysis}
\label{obs}
The 2006 December 26  and 2007 January 2 observations of M31 with XMM-Newton \citep{jan01} were conducted as part of the M31 extended survey program (PI W. Pietsch), which expands on the 2002 major axis survey to cover the entire optical D$_{25}$ region of M31 \citep[see ][]{stiele07}. In this work, we used the data from the European Photon Imaging Cameras (EPIC):  EPIC-pn \citep{stru01} and EPIC-MOS \citep{turn01}. We reconstructed the EPIC-pn and EPIC-MOS events files using the latest version of the XMM-Newton analysis package, SAS 7.1. A journal of observations for the three targets is presented in Table~\ref{journal}.

XMM-Newton observations are known to experience intervals of high background, so we screened these flares in the recommended manner. For the EPIC-pn, we created a 10--12 keV lightcurve with 100 s binning from the whole field of view, selecting only single events (PATTERN==0) from good pixels ( $\#$XMMEA$\_$EP); we rejected all intervals with $>$0.4 count s$^{-1}$. Similarly, we created lightcurves for each EPIC-MOS detector, for all  single events with energy $>$10 keV, with the equivalent EPIC-MOS filter ($\#$XMMEA$\_$EM). After checking that the flaring intervals were  the same in the EPIC-MOS, we filtered the EPIC-pn and EPIC-MOS events files on the EPIC-pn flare filter; resultant good time exposures are provided in Table~\ref{journal}.

For each source, we created a source extraction region and corresponding background region that was at least as big as the source region, on the same CCD, at a similar off-axis angle and contained no point sources { or extended emission}. For the EPIC-pn events, we accepted PATTERN$<$=4, and FLAG==0, while for EPIC-MOS events we accepted PATTERN$<$=12 and $\#$XMMEA$\_$EM. We created 0.3--10 keV source and background spectra, along with the appropriate response files. We also created synchronised 0.3--10 keV lightcurves for the EPIC-pn, EPIC-MOS1 and EPIC-MOS2 detectors \citep[see ][]{bko07, bs07}.  These lightcurves were then combined and their variability examined.

{ 
To obtain astrometrically-corrected positions we selected the optical
sources from the USNO-B1, 2MASS and Local Group Survey \citep{mas06} catalogues, and checked that only one optical source was in the
error circle of the corresponding X-ray source. We only accepted sources
correlating with globular clusters from the Revised Bologna Catalogue (V3.4) or with foreground stars, characterized by their optical
to X-ray flux ratio  \citep{mac88} and their hardness ratio
\citep[see][]{stiele08}. For sources selected from the USNO-B1
catalogue, we used proper-motion corrected positions. We then used the
SAS-task {\tt eposcorr} to derive the offset of the X-ray aspect solution.

}

\section{Results}
\label{res}

\subsection{Lightcurve analysis}

{
The background regions of XBo\thinspace 45, XBo\thinspace 135 and XBo\thinspace 375 had radii of 30$''$, 50$''$ and 40$''$ respectively, while  the source  regions had radii of 20$''$, 20$''$ and 40$''$; these source regions were chosen to optimally balance encircled energy, background subtraction and source crowding.   We therefore normalised the background lightcurves by the background to source area ratios.  The flare-free, normalised  background lightcurves were stable at $\sim$0.03 count s$^{-1}$ for all sources.

}

\begin{figure}[!t]
\resizebox{\hsize}{!}{\includegraphics[angle=0,scale=0.6]{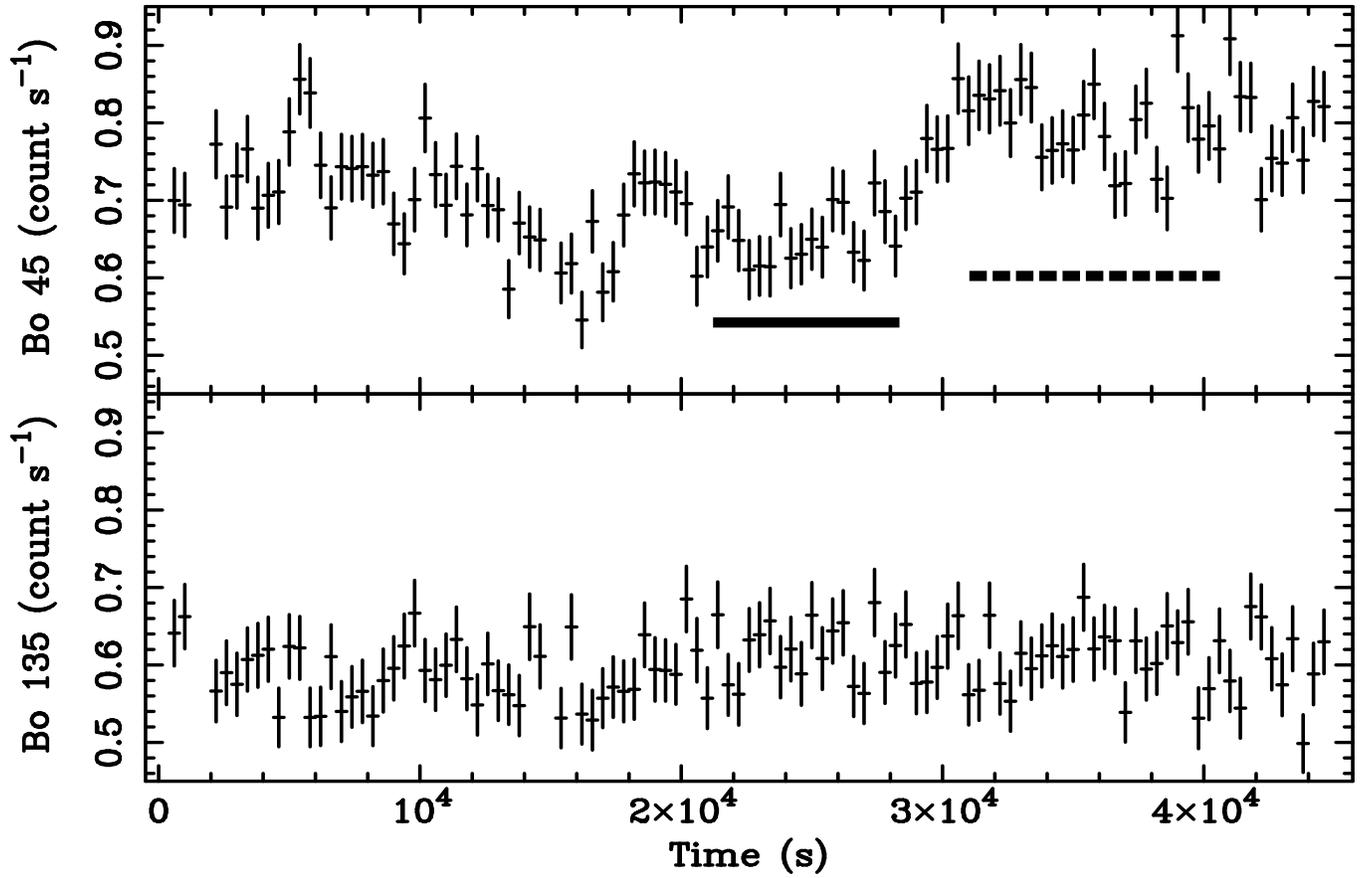}}
\caption{Combined EPIC-pn+EPIC-MOS 0.3--10 keV lightcurves of XBo\thinspace 45 and XBo\thinspace 135, binned to 400 s. The lightcurves have been background-subtracted and screened for flaring. The lightcurve of XBo\thinspace 45 is clearly variable; its r.m.s. variability is 9.2$\pm$0.9\%. However, XBo\thinspace 135 shows no significant variability. The solid and dashed lines in the XBo\thinspace 45 lightcurve indicate the intervals used for the low and high intensity spectra respectively, discussed in Section 5.3.\label{lcs}}
\end{figure}

\begin{figure}[!t]
\resizebox{\hsize}{!}{\includegraphics[angle=270,scale=0.6]{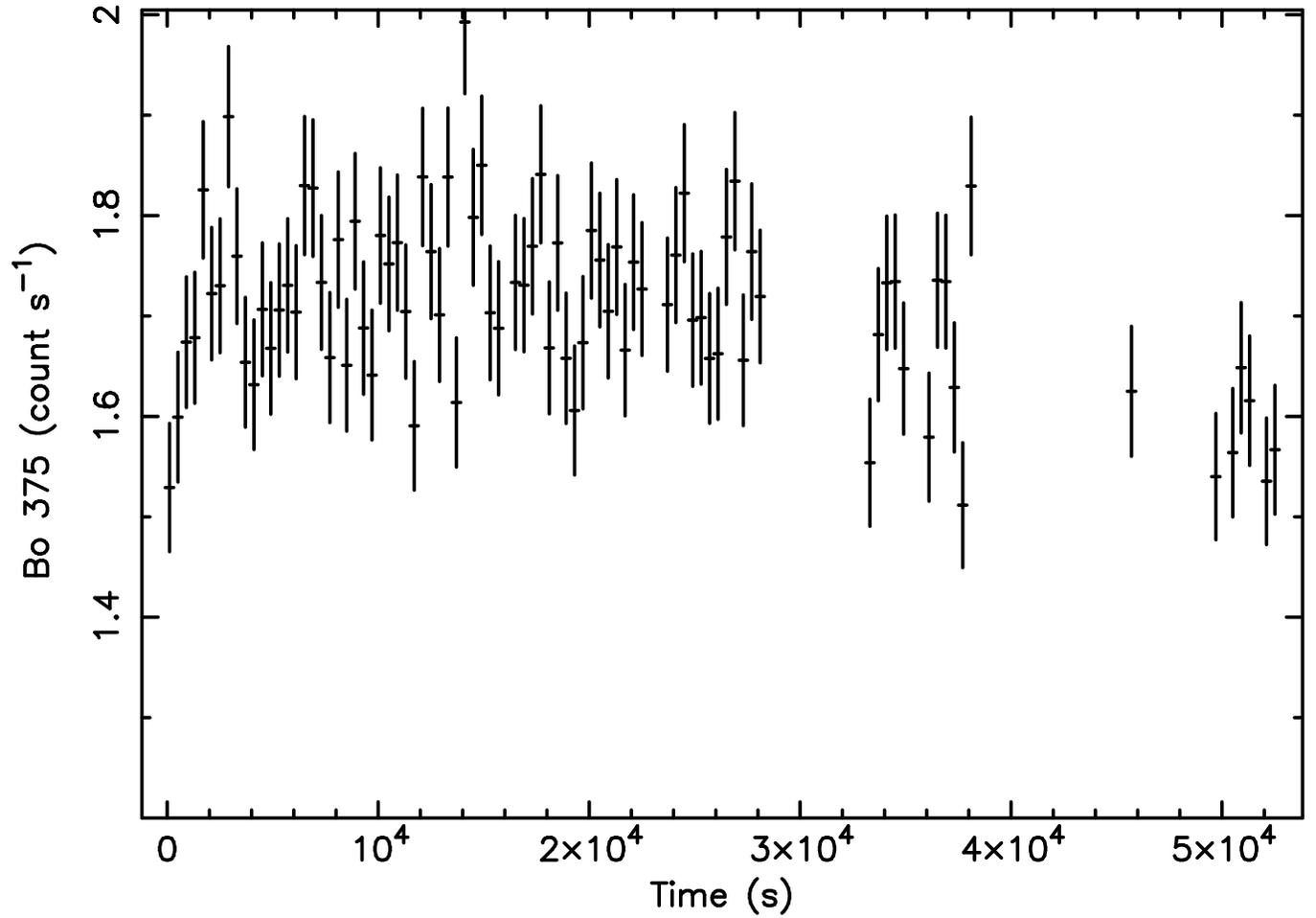}}
\caption{Combined EPIC-pn+EPIC-MOS lightcurve of XBo\thinspace 375 in the 0.3--10 keV band, binned to 400 s. Even though the count rate for XBo\thinspace 375 is 2--3 times higher than for XBo\thinspace 45 or XBo\thinspace 135, it is significantly less variable than XBo\thinspace 45.\label{lc375}}
\end{figure}

We present the 0.3--10 keV lightcurves of XBo\thinspace 45 and XBo\thinspace 135 in Fig.~\ref{lcs}, binned to 400 s.  XBo\thinspace 45 is clearly variable; the r.m.s. variability is 9.2$\pm$0.9\%, while the best fit line of constant intensity has a $\chi^2$/dof = 356/108.  Such variability is expected for low-state LMXBs, { with most of the power at frequencies $>$1 Hz; hence, the true variability of XBo\thinspace 45 is expected to be higher than $\sim$10\%}.  We analysed the power density spectrum of XBo\thinspace 45, in search of the characteristic broken power law PDS observed in low-state LMXBs. However, no significant power was detected in the PDS. This is likely to be due to faintness of the source. We note that XBo\thinspace 135 exhibited no detectable variability; the best fit line of constant intensity yielded $\chi^2$/dof = 115/108, and its r.m.s. variability was found to be 1.8$\pm$1.8\%. 
The  EPIC-pn+EPIC-MOS 0.3--10 keV lightcurve of XBo\thinspace 375 is shown in  in Fig.~\ref{lc375}. It is significantly less variable than XBo\thinspace 45 ($\chi^2$/dof = 153/86, 3.4$\pm$0.6\% variability) despite having more than twice the intensity. 

\subsection{Spectral analysis}

For each source we simultaneously fitted the EPIC-pn and EPIC-MOS spectra using XSPEC 11.3.21. Table~\ref{journal} provides the number of net source counts in each detector from each source.  For the models discussed below we assumed Solar abundances for our absorber \citep{ag89}. If we use the ISM abundances of \citet{wilm00}, and  the TBabs absorption model in XSPEC, then we obtain column densities $\sim$30\% higher, but similar emission parameters and luminosities. However, the resulting fits are slightly worse.

\subsubsection{Fitting a power law emission model}
\begin{figure}
\resizebox{\hsize}{!}{\includegraphics[angle=270,scale=0.6]{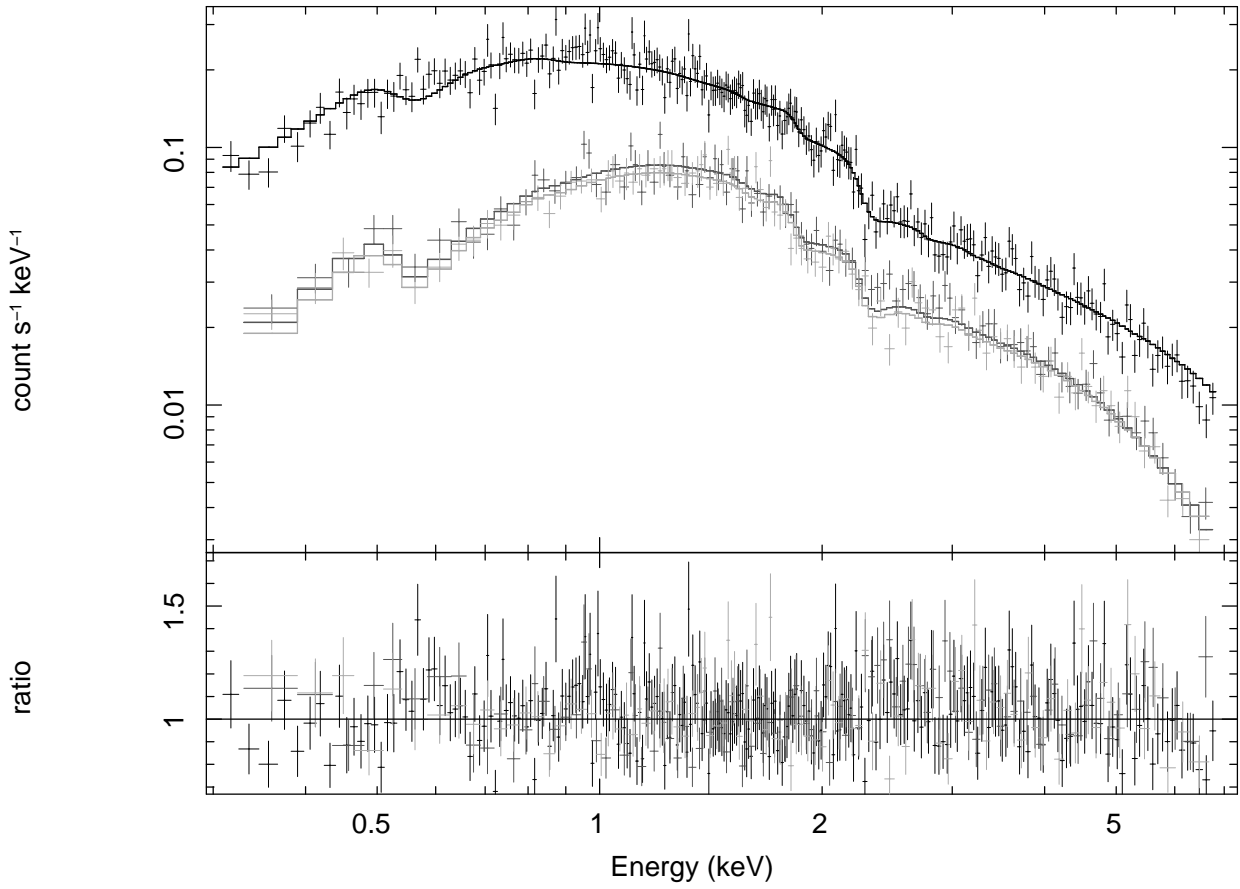}}
\caption{Best fit power law model to simultaneously fitted 0.3--7 keV EPIC-pn (black) and EPIC-MOS (light and dark grey) spectra of XBo\thinspace 45. The upper panel shows the log-scaled spectra, while the lower panel shows  the ratios of the observed to the expected flux for each energy bin}.\label{bh1spec}
\end{figure}

\begin{figure}
\resizebox{\hsize}{!}{\includegraphics[angle=270,scale=0.6]{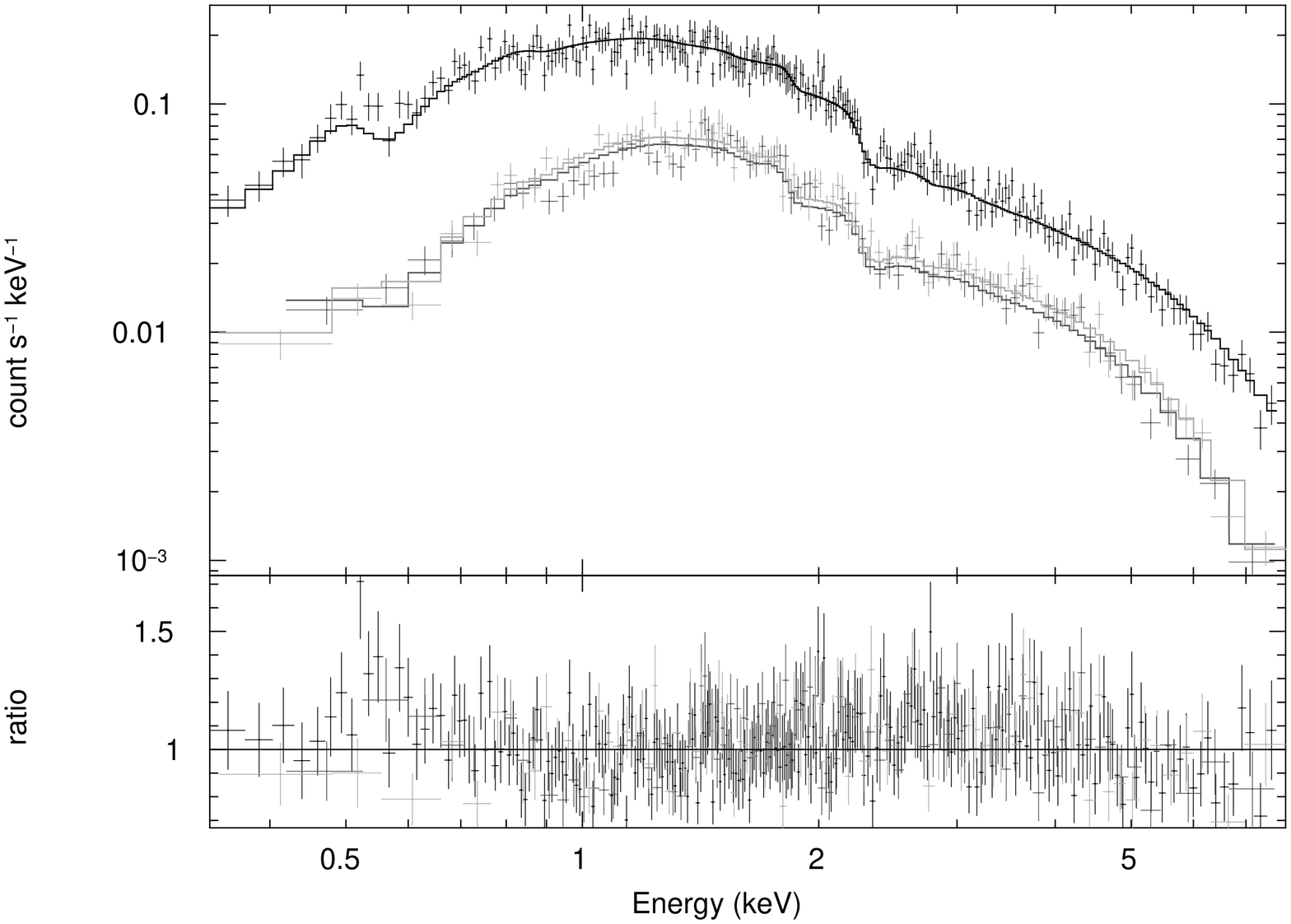}}
\caption{Best fit power law model to simultaneously fitted 0.3--7 keV EPIC-pn (black) and EPIC-MOS ( light and dark grey) spectra of XBo\thinspace 135. The upper panel shows the log-scaled spectra, while the lower panel shows ratios}.\label{bh2spec}
\end{figure}

 {  \citet{wsp88} found the spectra of low luminosity (i.e. low state) LMXBs to be well described by inverse Compton scattering of cool photons on hot electrons in a corona, and a power law  is a simplified representation of this model.}  Both XBo\thinspace 45 and XBo\thinspace 135 were well described by  an absorbed,  hard power law (see Table~\ref{spectab} for parameters). The best fit power law models to the pn and MOS spectra of XBo\thinspace 45  and XBo\thinspace 135 are presented in Fig.~\ref{bh1spec} and Fig.~\ref{bh2spec} respectively. The quoted uncertainties correspond to 90\% confidence limits. However, the best fit absorbed power law model to the XBo\thinspace 375 must be rejected.  

We also modeled the spectra of XBo\thinspace 45 and XBo\thinspace 135  with the {\sc comptt} spectral model \citep[see][]{tit94}. Our spectra could not sensibly constrain the temperature, favouring $\sim$50 keV.  

We calculated the luminosities of  XBo\thinspace 45 and XBo\thinspace 135 using the best fit power law emission models, and a distance of 784 kpc; uncertainties in the distance will be discussed below. The unabsorbed 0.3--10 keV luminosities of XBo\thinspace 45 and XBo\thinspace 135 were 2.46$\pm$0.09$\times$10$^{38}$ and 4.76$\pm$0.11$\times$10$^{38}$ erg s$^{-1}$ respectively.

\subsubsection{Adding a blackbody component}

\begin{figure}
\resizebox{\hsize}{!}{\includegraphics[angle=270,scale=0.6]{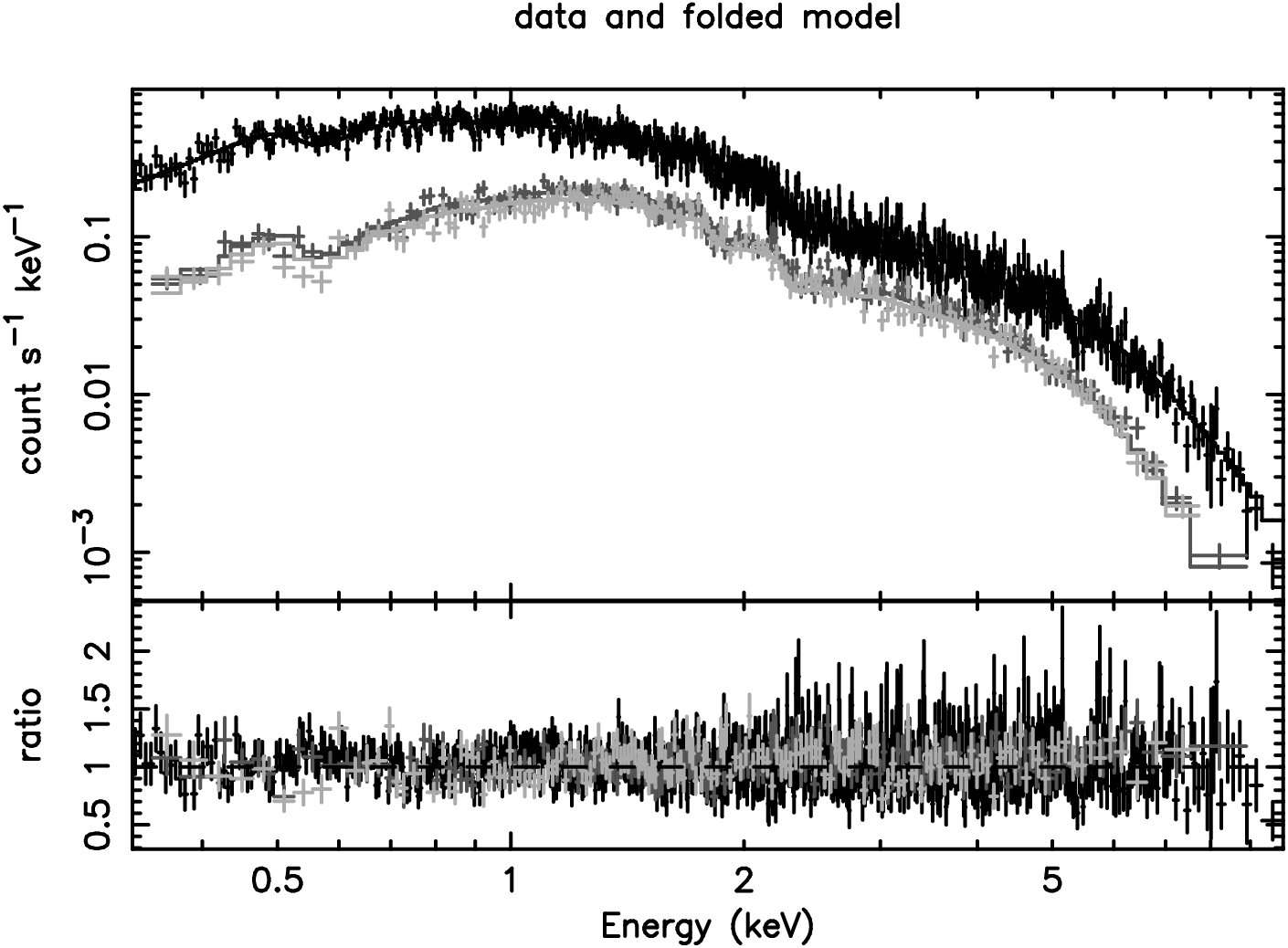}}
\caption{Best fit power law model to simultaneously fitted 0.3--7 keV EPIC-pn (black) and EPIC-MOS ( light and dark grey) spectra of XBo\thinspace 375. The upper panel shows the log-scaled spectra, while the lower panel shows ratios}.\label{375spec}
\end{figure}

A power law emisssion model is rejected for XBo\thinspace 375; however, adding a second, blackbody,  component  yields a good fit, shown in Table~\ref{spectab}.  The pn and MOS spectra of XBo\thinspace 375, along with this best fit model, are presented in Fig.~\ref{375spec}. The best fit, unabsorbed 0.3--10 keV luminosity for this source is  6.5$\pm 0.2 \times 10^{38}$ erg s$^{-1}$. The blackbody component contributes 14$\pm$4\%. We note that our values of k$T$ and $\Gamma$ are consistent with those found by \citet{dis02} for the ACIS-S spectrum; however, our preferred absorption is a factor $\sim$2 lower. We note that \citet{dis02} found the absorption to vary by a factor of $\sim$3 between ROSAT, and Chandra observations, so this is likely to be real.

 For XBo\thinspace 45, the fit is marginally improved by adding a blackbody component; however, the blackbody contribution to the flux could not be sensibly constrained. We infer from this that there is no significant blackbody contribution to the spectrum of XBo\thinspace 45.

Adding a blackbody component to the XBo\thinspace 135 emission spectrum improves the fit significantly. The best fit  BB+PO model is shown in Table ~\ref{spectab}. The 0.3--10 keV luminosity for this model is { 4.4$\pm$0.2$\times$ 10$^{38}$ erg s$^{-1}$}; the blackbody contributes 11$\pm$5\%.

The spectra of XBo\thinspace 375 had twice as many counts as the equivalent spectra of XBo\thinspace 45 or XBo\thinspace 135. Hence we created an  EPIC-pn spectrum for XBo\thinspace 375   with a similar number of source photons to the EPIC-pn spectra of XBo\thinspace 45 and XBo\thinspace 135, to see if the blackbody component was still necessary. The best fit power law yielded $\chi^2$/dof = 324/251 (g.f.p = 0.0013); hence,  a blackbody component was still required; the best fit BB+PO model yielded $\chi^2$/dof = 265/249 (g.f.p. = 0.22). These results show that the emission of XBo\thinspace 375 is significantly different from that of XBo\thinspace 45, where no blackbody is required.

We then modeled the spectra of each source with absorbed blackbody and disc blackbody  models, as might be expected for a BH LMXB in the high state. The best fit blackbody models yielded $\chi^2$/dof $>$4 for each source. The disc blackbody models fared better, but still must be rejected, with good fit probabilities ranging from 0.02 for XBo\thinspace 135 to 10$^{-7}$ for XBo\thinspace 375.  
Hence, we may securely reject thermally dominated emission for each source.

\subsubsection{Fitting low-metallicity absorbers}

\citet{dis02} investigated the effect on modelling XBo\thinspace 375 spectra of assuming  a metallicity appropriate to XBo\thinspace 375 (6\% Solar) for the absorber. Using the {\sc vphabs} model in XSPEC, they set the H and He abunances to cosmic values, and fixed the abundances of higher elements to 6\% Solar. Their modelling of the Chandra data  was improved ($\chi^2$/dof = 1.29 for 206 dof, c.f. 1.53 for 209 dof for a standard absorber); however, neither fit was formally acceptable (good fit probability $<$0.003). Neither ROSAT nor ASCA spectra were improved by low metallicity fits. 

We therefore modeled our XMM-Newton spectra of XBo\thinspace 45, XBo\thinspace 135 and XBo\thinspace 375, using metallicities of 0.11, 0.02 and 0.06 respectively, as obtained by \citet{bel95}. We present the best fit models for each source in Table~\ref{metal}. We see that the fits are considerably worse in each case, with only the BB+PO fit to XBo\thinspace 135 being a good fit. It is likely that the metallicites of the environs surrounding these sources  were enhanced by the supernovae that produced the NS/BH accretors, and therefore the GC metallicities are inappropriate.

\subsection{ The  spectral evolution of XBo\thinspace 45}

Of the three sources, only XBo\thinspace 45 exhibited significant time variability.
The observed intensity variation in XBo\thinspace 45 could be due to one of two phenomena: variation in the emission spectrum or variation in the absorbing material.   
To test these scenarios, we obtained two additional EPIC-pn spectra: one from a low intensity interval, and another from an interval of high intensity, represented  in Fig.~\ref{lcs} by a solid line and a dashed line respectively. We applied the same response and background files to these spectra as to the original XBo\thinspace 45 spectrum. The low intensity spectrum contained $\sim$3000 source counts, for a mean intensity of 0.424$\pm$0.008 count s$^{-1}$, while the high intensity spectrum contained $\sim$4500 source counts, with a mean intensity of 0.512$\pm$0.008 count s$^{-1}$. 

Simultaneously fitting the low and high intensity spectra with $\Gamma$ and normalisation linked but free to vary, and  with the absorption ($N_{\rm H}$) free to vary yielded an unacceptable best fit. We then { linked} $N_{\rm H}$ and $\Gamma$, varying only the normalisation; this produced an acceptable fit. Finally, we freely fitted each spectrum, and found that $N_{\rm H}$ and $\Gamma$ were consistent within 90\% confidence limits for the two spectra. The best fits for each of these models are presented in Table~\ref{t2}. It is clear that the variation is intrinsic to the X-ray source, rather than the absorber; such variation is characteristic of low state LMXBs \citep{vdk94}.

{\subsection{Are the X-ray sources associated with the GCs?}

Mass segregation is thought to concentrate binaries, which are amongst the heaviest objects in globular clusters, to the centre of the cluster \citep[e.g.][]{mh97}.
The X-ray sources are each located $\sim$1$''$ from their respective GC positions, with positional errors of $\sim$1$''$.  Of the three GCs, only Bo\thinspace 45 has sufficient quality data to measure its structure. \citet{barm07} used HST observations of  Bo 45 to model its spatial characteristics, and found a core radius of 0.3$''$ and a half-mass radius of 1$''$. It is therefore impossible to confirm whether the X-ray sources are located in the central regions of the GCs.

 Following \citet{mac07}, we first calculated the probability that the X-ray sources have GC associations purely through chance, and then calculated the probability that these sources are active galacic nuclei (AGN) that are coincident with the GCs.  

Both XBo\thinspace 45 and XBo\thinspace 135 were observed in the NN1 region (Table 1).
 Assuming a circular field of view with 15$'$ radius, we searched the literature for GCs and GC candidates within the NN1 observation, finding 16 GCs and 51 GC candidates. Hence the GC spatial density $\le$3$\times$10$^{-5}$ per square arcsec. We found 30 X-ray sources with $>$100 photons in the EPIC-pn camera; therefore the probability of finding a GC within 1$''$ of a bright X-ray sources is $\le$9$\times$10$^{-4}$, and the probability of two chance associations is $\le$8$\times$10$^{-7}$.

XBo\thinspace 375 was observed in the NS2 region (Table 1). We found 19 GCs and 39 GC candidates, giving a spatial density $\le$2$\times$10$^{-5}$. There were 20 sources with $>$100 EPIC-pn counts, and the probability of a GC coinciding with one of these by chance is $\le$4$\times$10$^{-4}$.

We estimated the probability that the X-ray sources were AGN by utilising the work of \citet{mor03}, who have modeled the AGN flux distributions in the 1--2 keV and 2--10 keV bands. We found 0.33$^{+0.36}_{-0.2}$ AGN per square degree with 1--2 keV fluxes as high as the 1--2 keV flux of XBo\thinspace 45 (4.4$\times$10$^{-13}$ erg cm$^{-2}$ s$^{-1}$), and 0.08$^{+0.14}_{-0.05}$ AGN per square degree with 2--10 keV fluxes as high for XBo\thinspace 45  (2.3$\times$10$^{12}$   erg cm$^{-2}$ s$^{-1}$). Even if we assume that none of these 2--10 keV sources are included in the 1--2 keV sources, and take the upper limits, then the spatial density of AGN as bright as XBo\thinspace 45 is $\le$7$\times$10$^{-8}$ per square arcsec; the probability of such an AGN coinciding with a GC is $\le$5$\times$10$^{-6}$. Furthermore, XBo\thinspace 135 and XBo\thinspace 375 are more luminous than XBo\thinspace 45, and they are even less likely to be AGN. 

We also note that any AGN, or indeed foreground star, would be very likely to distort the optical colours of the cluster.  
We therefore conclude that the X-ray sources are very probably associated with the GCs. 
}

\section{Discussion and conclusions}
\label{dis}
We have examined the emission spectra and time variability of three X-ray sources associated with GCs in M31, using the 2006 December 26 and 2007 January 2 XMM-Newton observations. 
The emission  of XBo\thinspace 45 is well described by a pure power law with photon index $\sim$1.4, and is highly variable. This is consistent with a NS or BH LMXB in the low state \citep{vdk94}, but is not consistent with a  BH LMXB in the high state or steep power law state \citep[and references within]{mr03}, or  a NS LMXB emitting at $>$10$^{38}$ erg s$^{-1}$ \citep{wsp88,cbc01,bcb03}.

{ \citet{sg98} calculated a distance to M31 of 784 kpc, $\pm$13 kpc of statistical error, $\pm$17 kpc of systematic error. Combining these distance uncertainties adds further uncertainties in the source luminosities of $\pm$5\%. Therefore, the 0.3--10 keV luminosity range for XBo\thinspace  45 is 2.5$\pm$0.2$\times$10$^{38}$ erg s$^{-1}$, or 140$\pm$10\% of the Eddington limit for a 1.4 M$_{\odot}$ neutron star primary. However, several LMXBs have been found to contain neutron stars with mass as high as $\sim$2.1 M$_{\odot}$ \citep[see e.g.][]{nice05,ozel06}; XBo\thinspace 45 has a 0.3--10 keV luminosity of $\sim$80\% Eddington for such systems. Since \citet{glad07} showed that transitions in neutron star LMXBs occur at $\la$10\% Eddington, we consider XBo\thinspace 45 to exhibit low state behaviour at an apparent luminosity too high for a neutron star. We therefore identify the accretor in XBo\thinspace 45 as a BH candidate. 
 We note that XBo\thinspace  45  has been observed several times by the Einstein and ROSAT observatories over the last $\sim30$ years, varying in luminosity only by a factor of $\sim$2. This behaviour is consistent with that predicted for a GC BH binary formed by tidal capture \citep{kal04}.

In contrast to XBo\thinspace 45,} the observed two component emission  of XBo\thinspace 375 is consistent with a bright NS LMXB, but not a low state NS or BH LMXB, nor a BH in the high or steep power law states. Hence, we classify XBo\thinspace 375 as a NS LMXB.  Finally, the emission spectrum of XBo\thinspace 135 is consistent with a pure power law with photon index $\sim$1.6, but the fit is significantly improved by adding a blackbody component. Hence, XBo\thinspace 135 is consistent with a NS or BH LMXB, and deeper observation is required for further classification. 

{ 
Of the thirteen bright X-ray sources in Galactic GCs, twelve have exhibited X-ray bursts, confiming their natures as NS LMXBs \citep[see ][ for a review]{vl06}. \citet{ph05} conducted a survey of  XMM-Newton observations of X-ray sources associated with M31 GCs, looking for X-ray bursts. They found simultaneous pn and MOS detections of bursts in two sources, and several burst candidates that were detected in the pn only. Such bursts would be identifiable in the lightcurves of our target sources. However, X-ray bursts are thought to be forbidden at luminosities $>$ 50\% Eddington \citep{lpt93}, hence we do not expect bursts from XBo\thinspace 135 or XBo\thinspace 375.
}

XBo\thinspace 45 exhibits low-state behaviour at luminosities $\sim$10 times higher than expected for 1.4 M$_{\odot}$ NS LMXBs; XBo\thinspace 135 may also. A composite of two or three bright X-ray sources would likely result in a different spectrum to that observed in XBo\thinspace 45; hence, if XBo\thinspace 45 were a composite, it would be more likely to be made from { $\sim$10} NS LMXBs in the low state. \citet{dis02} calculated the probability for a GC to contain multiple bright X-ray sources: they argue that if $p$ is the probability of finding one  bright X-ray source in a GC, then Poisson statistics dictate that probability for two bright X-ray sources should be  $p$/2, while the probability for 3 bright X-ray sources should be  $p^{2}$/6. For a $p$ $\sim$0.1--0.2, \citet{dis02}  predict 3--5 GCs with 2 X-ray sources and less than 1 with 3 X-ray sources in M31. It is therefore very unlikely that XBo\thinspace 45 combines the emission of $\sim$10  low-state neutron star LMXBs.

It is however possible that the emission from XBo\thinspace 45  is anisotropic. It exhibits low-state spectra at luminosities $\sim$10--20 times higher than expected for neutron star LMXBs; hence it could simply be beamed by a factor of 10--20. In this case, the emission would be restricted to a small solid angle; one might expect to observe such beaming in $\sim$5--10\% of randomly oriented systems. \citet{tp04} modeled the spectra of 43 M31 GCs. If we loosely class a low-state spectrum as a power law with $\Gamma$ $<$1.7, then \citet{tp04} found 20 GCs consistent with low-state spectra. Five of those,  XBo\thinspace 5, XBo\thinspace 82, XBo\thinspace 135,  XBo\thinspace 153 and XBo\thinspace 386 exhibited apparent luminosities $>$10$^{38}$ erg s$^{-1}$. If we include XBo\thinspace 45, then 6 out of   21 GC systems consistent with low-state spectra (including XBo\thinspace 45), have measured luminosities $>$10$^{38}$, i.e. $\sim$30\%, a significantly larger fraction than expected from beaming.

We note that the host cluster Bo\thinspace 45, which contains a BH LMXB candidate, is significantly larger than the cluster Bo\thinspace 375 { (see Section 3.2)}, which we think contains a NS LMXB. \citet{dis02} describe Bo\thinspace 375 as not at all unusual, with parameters close to the median of M31 GCs. This suggests that Bo\thinspace 45 (and also Bo\thinspace 135) are particularly  massive, and may therefore be more prone to forming BH LMXBs. Therefore we conclude that, unlike the
Milky Way, at least one GC in M31 is likely to contain a black hole binary.

\acknowledgements
This work is based on observations with XMM-Newton, an ESA science mission with  instruments and contributions directly funded by ESA member states and the USA (NASA).
Astrophysics research at the Open University is funded by a STFC rolling grant. HS acknowledges support by the Bundesministerium f\"ur Wirtschaft und
Technologie/Deutsches Zentrum f\"ur Luft- und Raumfahrt  (BMWI/DLR, FKZ
50 OR 0405). GS is supported through DLR (FKZ 50 OR 0405). { This publication makes use of the USNOFS Image and Catalogue Archive
operated by the United States Naval Observatory, Flagstaff Station
(http://www.nofs.navy.mil/data/fchpix/) and
of data products from the Two Micron All Sky Survey,
which is a joint project of the University of Massachusetts and the
Infrared
Processing and Analysis Center/California Institute of Technology,
funded by
the National Aeronautics and Space Administration and the National Science
Foundation}.

\clearpage

\begin{deluxetable}{cccccccc}
\tabletypesize{\scriptsize}
\tablecaption{Journal of observations. For each object we give the position, XMM-Newton observation, good time exposure, and number of net source photons in the EPIC-pn, EPIC-MOS1 and EPIC-MOS2 spectra.\label{journal}}
\tablewidth{0pt}
\tablehead{\colhead{Source} & \colhead{ X-ray Position} & \colhead{Observation} & \colhead{Date} & Exp&  \colhead{pn cnt} & \colhead{MOS1 cnt} & MOS2 cnt} 
\startdata
XBo\thinspace 45 &  00$^{\rm h}$41$^{\rm m}$ 43.19$^{\rm s}$ +41$^{\circ}$34$'$20.1$''$ & NN1 (00$^{\rm h}$41$^{\rm m}$52.8$^{\rm s}$ $+41^{\circ}$36$'$36.0$''$) & 2006-12-26 & 40 ks & 16894 & 7416 & 6953\\

XBo\thinspace 135 &00$^{\rm h}$42$^{\rm m}$52.00$^{\rm s}$ +41$^{\circ}$31$'$09.7$''$ & NN1 & 2006-12-26 & 40 ks &15104 & 5560 & 6090\\ 
XBo\thinspace 375 & 00$^{\rm h}$45$^{\rm m}$45.54$^{\rm s}$ +41$^{\circ}$39$'$42.6$''$& NS2 (00$^{\rm h}$45$^{\rm m}$50.4$^{\rm s}$ +41$^{\circ}$30$'$44.7$''$) & 2007-01-02  & 32 ks& 31946 & 15686 & 13798 \\
\enddata
\end{deluxetable}

\clearpage

\begin{deluxetable}{cccccccc}
\tabletypesize{\scriptsize}
\tablecaption{ Best fit spectral models for fitting 0.3--7.0 keV EPIC-pn and EPIC-MOS spectra from XBo\thinspace 45, XBo\thinspace 135 and XBo\thinspace 375. The models are absorbed power law (PO), and absorbed blackbody+power law (BB+PO). For each model  we give the column density ($n_{\rm H}$/10$^{21}$ cm$^{-2}$),  blackbody temperature (k$T$/keV), photon index ($\Gamma$), constant of normalisation for EPIC-MOS1 and EPIC-MOS2 ($N_{\rm M1}$ and $N_{\rm M2}$), $\chi^2$/dof [good fit probability] and 0.3--10 keV unabsorbed flux. Numbers in parentheses indicate the 90\% uncertainty in the final digits.\label{spectab}}
\tablewidth{0pt}
\tablehead{\colhead{Source} & \colhead{$n_{\rm H}$/10$^{21}$ cm$^{-2}$} & \colhead{k$T$/keV}& \colhead{$\Gamma$} & \colhead{$N_{\rm M1}$} & \colhead{$N_{\rm M2}$} & \colhead{$\chi^2$/dof [gfp]} & \colhead{$F_{0.3-10}$/10$^{-12}$ erg cm$^{-2}$ s$^{-1}$}} 
\startdata
XBo\thinspace 45 PO &  1.41(11) & $\dots$ & 1.45(4) & 1.06(4)& 1.03(4) & 517/487 [0.17] & 3.34(12)\\
XBo\thinspace 45 BB+PO & 1.46(19) & 1.23(19) & 1.57(6) & 1.06(3) & 1.03(2) & 501/485 [0.30] & 3.2(2)\\
XBo\thinspace 135 PO & 2.76(12) & $\dots$ &  1.56(3) & 1.12(3) & 1.06(3)  & 467/435 [0.14] & 6.45(15)\\ 
XBo\thinspace 135 BB+PO & 2.3(3) & 0.8(2) & 1.54(14) & 1.12(3) & 1.06(3) & 413/433 [0.75] & 6.0(3)\\
XBo\thinspace 375 PO & 1.52(6) & $\dots$ & 1.64(2) & 1.15(2) & 1.07(2) & 1180/1032 [9E-4] & 9.17(17)\\
XBo\thinspace 375 BB+PO & 1.41(11) & 0.90(10) & 1.73(18) & 1.15(2) & 1.07(2) & 1110/1030 [0.19] & 8.8(3)\\
\enddata
\end{deluxetable}

\clearpage

\begin{deluxetable}{cccccccc}
\tabletypesize{\scriptsize}
\tablecaption{ Best fit models  for fitting 0.3--7.0 keV EPIC-pn and EPIC-MOS spectra from XBo\thinspace 45,  XBo\thinspace 135 and XBo\thinspace 375 using observed GC metallicities. For each source we give the column density ($n_{\rm H}$/10$^{21}$ cm$^{-2}$),  temperature (k$T$/keV), photon index ($\Gamma$), constant of normalisation for EPIC-MOS1 and EPIC-MOS2 ($N_{\rm M1}$ and $N_{\rm M2}$),  and  $\chi^2$/dof [good fit probability]. Numbers in parentheses indicate the 90\% uncertainty in the final digits.\label{metal}}
\tablewidth{0pt}
\tablehead{\colhead{Source} & \colhead{$n_{\rm H}$/10$^{21}$ cm$^{-2}$} & \colhead{k$T$}/keV & \colhead{$\Gamma$} & \colhead{$N_{\rm M1}$} & \colhead{$N_{\rm M2}$} & \colhead{$\chi^2$/dof [gfp]} } 
\startdata
XBo\thinspace 45 PO & 0.188(12) & $\dots$ & 1.33(2) & 1.06(2) & 1.05(2) & 567/487 [7E-3]\\
XBo\thinspace 135 PO & 0.58(4) & $\dots$ & 1.39(2) & 1.13(3) & 1.06(3) & 740/435 [3E-18]\\
XBo\thinspace 135 BB+PO & 0.24(4) & 0.69(2) & 1.05(5) & 1.13(3) & 1.07(3) & 436/433 [0.45] & \\
XBo\thinspace 375 BB+PO & 0.131(14) & 0.69(4) & 1.42(3) & 1.15(2) & 1.07(2) & 1110/1030[0.04]\\
\enddata
\end{deluxetable}

\clearpage
\begin{deluxetable}{ccccccccccc}
\tabletypesize{\scriptsize}
\tablecaption{Results from simultaneously fitting 0.3--7 keV EPIC-pn spectra from the high and low intensity intervals in the lightcurve of XBo\thinspace 45. For each model, we present absorption ($N_{\rm H}$ / 10$^{21}$), photon index ($\Gamma$), best fit $\chi^2$/dof and goodness of fit. Numbers in paranthese represent 90\% uncertainties in the last digits.\label{t2}}
\tablewidth{0pt}
\tablehead{\colhead{} &\multicolumn{3}{c}{Variable absorption } & \multicolumn{3}{c}{Variable normalisation} & \multicolumn{3}{c}{Free fitting}\\
\colhead{} & \colhead{$N_{\rm H}$/10$^{21}$ cm$^{-2}$} & \colhead{$\Gamma$} & \colhead{$\chi^2$/dof [gfp]} &  \colhead{$N_{\rm H}$/10$^{21}$ cm$^{-2}$} & \colhead{$\Gamma$} & \colhead{$\chi^2$/dof [gfp]} &  \colhead{$N_{\rm H}$/10$^{21}$} & \colhead{$\Gamma$} & \colhead{$\chi^2$/dof [gfp]} } 
\startdata
High& 1.4(2) & 1.51(8) & 193/129 [1.9e-4] & 1.5(2) & 1.50(8) & 140/129 [0.23] & 1.5(3) & 1.45(10) & 125/127 [0.54]\\
Low & 1.5(3) & 1.51(8) & & 1.5(2) & 1.50(8) & & 1.4(4) & 1.57(15) & \\ 
\enddata
\end{deluxetable}

\clearpage

\clearpage
 
\clearpage

\clearpage

\clearpage


\begin{thebibliography}{59}
\expandafter\ifx\csname natexlab\endcsname\relax\def\natexlab#1{#1}\fi

\bibitem[{{Anders} \& {Grevesse}(1989)}]{ag89}
{Anders}, E. \& {Grevesse}, N. 1989, \gca, 53, 197

\bibitem[{{Barmby} {et~al.}(2007){Barmby}, {McLaughlin}, {Harris}, {Harris}, \&
  {Forbes}}]{barm07}
{Barmby}, P., {McLaughlin}, D.~E., {Harris}, W.~E., {Harris}, G.~L.~H., \&
  {Forbes}, D.~A. 2007, \aj, 133, 2764

\bibitem[{{Barnard} {et~al.}(2003{\natexlab{a}}){Barnard}, {Church}, \& {Ba\l
  uci\'{n}ska-Church}}]{bcb03}
{Barnard}, R., {Church}, M.~J., \& {Ba\l uci\'{n}ska-Church}, M.
  2003{\natexlab{a}}, A\&A, 405, 237

\bibitem[{{Barnard} {et~al.}(2004){Barnard}, {Kolb}, \& {Osborne}}]{bko04}
{Barnard}, R., {Kolb}, U., \& {Osborne}, J.~P. 2004, A\&A, 423, 147

\bibitem[{{Barnard} {et~al.}(2007{\natexlab{a}}){Barnard}, {Kolb}, \&
  {Osborne}}]{bko07}
{Barnard}, R., {Kolb}, U.~C., \& {Osborne}, J.~P. 2007{\natexlab{a}}, \aap,
  469, 873

\bibitem[{{Barnard} {et~al.}(2003{\natexlab{b}}){Barnard}, {Osborne}, {Kolb},
  \& {Borozdin}}]{bok03}
{Barnard}, R., {Osborne}, J.~P., {Kolb}, U., \& {Borozdin}, K.~N.
  2003{\natexlab{b}}, A\&A, 405, 505

\bibitem[{{Barnard} {et~al.}(2007{\natexlab{b}}){Barnard}, {Trudolyubov},
  {Kolb}, {Haswell}, {Osborne}, \& {Priedhorsky}}]{bs07}
{Barnard}, R., {Trudolyubov}, S., {Kolb}, U.~C., {et~al.} 2007{\natexlab{b}},
  A\&A, 469, 875

\bibitem[{{Bellazzini} {et~al.}(1995){Bellazzini}, {Pasquali}, {Federici},
  {Ferraro}, \& {Pecci}}]{bel95}
{Bellazzini}, M., {Pasquali}, A., {Federici}, L., {Ferraro}, F.~R., \& {Pecci},
  F.~F. 1995, \apj, 439, 687

\bibitem[{{Church} \& {Ba\l uci{\' n}ska-Church}(2001)}]{cbc01}
{Church}, M.~J. \& {Ba\l uci{\' n}ska-Church}, M. 2001, A\&A, 369, 915

\bibitem[{{Church} \& {Ba\l uci\'{n}ska-Church}(1995)}]{cbc95}
{Church}, M.~J. \& {Ba\l uci\'{n}ska-Church}, M. 1995, A\&A, 300, 441

\bibitem[{{Di Stefano} {et~al.}(2002){Di Stefano}, {Kong}, {Garcia}, {Barmby},
  {Greiner}, {Murray}, \& {Primini}}]{dis02}
{Di Stefano}, R., {Kong}, A.~K.~H., {Garcia}, M.~R., {et~al.} 2002, ApJ, 570,
  618

\bibitem[{{Galleti} {et~al.}(2007){Galleti}, {Bellazzini}, {Federici},
  {Buzzoni}, \& {Fusi Pecci}}]{gal07}
{Galleti}, S., {Bellazzini}, M., {Federici}, L., {Buzzoni}, A., \& {Fusi
  Pecci}, F. 2007, \aap, 471, 127

\bibitem[{{Galleti} {et~al.}(2005){Galleti}, {Bellazzini}, {Federici}, \& {Fusi
  Pecci}}]{gal05}
{Galleti}, S., {Bellazzini}, M., {Federici}, L., \& {Fusi Pecci}, F. 2005,
  \aap, 436, 535

\bibitem[{{Galleti} {et~al.}(2006){Galleti}, {Federici}, {Bellazzini},
  {Buzzoni}, \& {Fusi Pecci}}]{gal06}
{Galleti}, S., {Federici}, L., {Bellazzini}, M., {Buzzoni}, A., \& {Fusi
  Pecci}, F. 2006, \aap, 456, 985

\bibitem[{{Galleti} {et~al.}(2004){Galleti}, {Federici}, {Bellazzini}, {Fusi
  Pecci}, \& {Macrina}}]{gall04}
{Galleti}, S., {Federici}, L., {Bellazzini}, M., {Fusi Pecci}, F., \&
  {Macrina}, S. 2004, \aap, 416, 917

\bibitem[{{Gladstone} {et~al.}(2007){Gladstone}, {Done}, \&
  {Gierli{\'n}ski}}]{glad07}
{Gladstone}, J., {Done}, C., \& {Gierli{\'n}ski}, M. 2007, MNRAS, 378, 13

\bibitem[{{in't Zand} {et~al.}(2004){in't Zand}, {Verbunt}, {Heise}, {Bazzano},
  {Cocchi}, {Cornelisse}, {Kuulkers}, {Natalucci}, \& {Ubertini}}]{int04}
{in't Zand}, J., {Verbunt}, F., {Heise}, J., {et~al.} 2004, {To appear in "The
  Restless High-Energy Universe" (2nd BeppoSAX Symposium), eds. E.P.J. van den
  Heuvel, J.J.M. in 't Zand \& R.A.M.J. Wijers, Nucl. Instrum. Meth. B Suppl.
  Ser, astro-ph/0403120}

\bibitem[{{Jansen} {et~al.}(2001){Jansen}, {Lumb}, {Altieri}, {Clavel}, {Ehle},
  {Erd}, {Gabriel}, {Guainazzi}, {Gondoin}, {Much}, {Munoz}, {Santos},
  {Schartel}, {Texier}, \& {Vacanti}}]{jan01}
{Jansen}, F., {Lumb}, D., {Altieri}, B., {et~al.} 2001, A\&A, 365, L1

\bibitem[{{Kaaret}(2002)}]{kaa02}
{Kaaret}, P. 2002, ApJ, 578, 114

\bibitem[{{Kalogera} {et~al.}(2004){Kalogera}, {King}, \& {Rasio}}]{kal04}
{Kalogera}, V., {King}, A.~R., \& {Rasio}, F.~A. 2004, ApJL, 601, L171

\bibitem[{{Kundu} {et~al.}(2008){Kundu}, {Zepf}, \& {Maccarone}}]{kun08}
{Kundu}, A., {Zepf}, S.~E., \& {Maccarone}, T.~J. 2008, ArXiv e-prints, 801

\bibitem[{{Lampton} {et~al.}(1976){Lampton}, {Margon}, \& {Bowyer}}]{lam76}
{Lampton}, M., {Margon}, B., \& {Bowyer}, S. 1976, \apj, 208, 177

\bibitem[{{Lewin} {et~al.}(1993){Lewin}, {van Paradijs}, \& {Taam}}]{lpt93}
{Lewin}, W.~H.~G., {van Paradijs}, J., \& {Taam}, R.~E. 1993, Space Science
  Reviews, 62, 223

\bibitem[{{Maccacaro} {et~al.}(1988){Maccacaro}, {Gioia}, {Wolter}, {Zamorani},
  \& {Stocke}}]{mac88}
{Maccacaro}, T., {Gioia}, I.~M., {Wolter}, A., {Zamorani}, G., \& {Stocke},
  J.~T. 1988, \apj, 326, 680

\bibitem[{{Maccarone} {et~al.}(2007){Maccarone}, {Kundu}, {Zepf}, \&
  {Rhode}}]{mac07}
{Maccarone}, T.~J., {Kundu}, A., {Zepf}, S.~E., \& {Rhode}, K.~L. 2007, \nat,
  445, 183

\bibitem[{{Massey} {et~al.}(2006){Massey}, {Olsen}, {Hodge}, {Strong},
  {Jacoby}, {Schlingman}, \& {Smith}}]{mas06}
{Massey}, P., {Olsen}, K.~A.~G., {Hodge}, P.~W., {et~al.} 2006, \aj, 131, 2478

\bibitem[{{McClintock} \& {Remillard}(2006)}]{mr03}
{McClintock}, J.~E. \& {Remillard}, R.~A. 2006, Compact stellar X-ray sources,
  157

\bibitem[{{Meylan} \& {Heggie}(1997)}]{mh97}
{Meylan}, G. \& {Heggie}, D.~C. 1997, \aapr, 8, 1

\bibitem[{{Miyamoto} {et~al.}(1992){Miyamoto}, {Kitamoto}, {Iga}, {Negoro}, \&
  {Terada}}]{miy92}
{Miyamoto}, S., {Kitamoto}, S., {Iga}, S., {Negoro}, H., \& {Terada}, K. 1992,
  ApJL, 391, L21

\bibitem[{{Moretti} {et~al.}(2003){Moretti}, {Campana}, {Lazzati}, \&
  {Tagliaferri}}]{mor03}
{Moretti}, A., {Campana}, S., {Lazzati}, D., \& {Tagliaferri}, G. 2003, ApJ,
  588, 696

\bibitem[{{Nice} {et~al.}(2005){Nice}, {Splaver}, {Stairs}, {L{\"o}hmer},
  {Jessner}, {Kramer}, \& {Cordes}}]{nice05}
{Nice}, D.~J., {Splaver}, E.~M., {Stairs}, I.~H., {et~al.} 2005, \apj, 634,
  1242

\bibitem[{{Oosterbroek} {et~al.}(1997){Oosterbroek}, {van der Klis}, {van
  Paradijs}, {Vaughan}, {Rutledge}, {Lewin}, {Tanaka}, {Nagase}, {Dotani},
  {Mitsuda}, \& {Miyamoto}}]{oost97}
{Oosterbroek}, T., {van der Klis}, M., {van Paradijs}, J., {et~al.} 1997, A\&A,
  321, 776

\bibitem[{{{\"O}zel}(2006)}]{ozel06}
{{\"O}zel}, F. 2006, \nat, 441, 1115

\bibitem[{{Pietsch} {et~al.}(2005){Pietsch}, {Freyberg}, \& {Haberl}}]{pfh05}
{Pietsch}, W., {Freyberg}, M., \& {Haberl}, F. 2005, \aap, 434, 483

\bibitem[{{Pietsch} \& {Haberl}(2005)}]{ph05}
{Pietsch}, W. \& {Haberl}, F. 2005, \aap, 430, L45

\bibitem[{{Rich} {et~al.}(2005){Rich}, {Corsi}, {Cacciari}, {Federici}, {Fusi
  Pecci}, {Djorgovski}, \& {Freedman}}]{ric05}
{Rich}, R.~M., {Corsi}, C.~E., {Cacciari}, C., {et~al.} 2005, \aj, 129, 2670

\bibitem[{{Sarajedini} {et~al.}(2007){Sarajedini}, {Barker}, {Geisler},
  {Harding}, \& {Schommer}}]{sar07}
{Sarajedini}, A., {Barker}, M.~K., {Geisler}, D., {Harding}, P., \& {Schommer},
  R. 2007, \aj, 133, 290

\bibitem[{{Shahbaz} {et~al.}(1994){Shahbaz}, {Ringwald}, {Bunn}, {Naylor},
  {Charles}, \& {Casares}}]{shab94}
{Shahbaz}, T., {Ringwald}, F.~A., {Bunn}, J.~C., {et~al.} 1994, MNRAS, 271, L10

\bibitem[{{Shaw Greening} {et~al.}(2008){Shaw Greening}, {Barnard}, {Kolb},
  {Tonkin}, \& {Osborne}}]{lsg08}
{Shaw Greening}, L., {Barnard}, R., {Kolb}, U., {Tonkin}, C., \& {Osborne},
  J.~P. 2008, ArXiv e-prints, 'A\&A accepted'

\bibitem[{{Shirey} {et~al.}(2001){Shirey}, {Soria}, {Borozdin}, {Osborne},
  {Tiengo}, {Guainazzi}, {Hayter}, {La Palombara}, {Mason}, {Molendi},
  {Paerels}, {Pietsch}, {Priedhorsky}, {Read}, {Watson}, \& {West}}]{shi01}
{Shirey}, R., {Soria}, R., {Borozdin}, K., {et~al.} 2001, A\&A, 365, L195

\bibitem[{{Stanek} \& {Garnavich}(1998)}]{sg98}
{Stanek}, K.~Z. \& {Garnavich}, P.~M. 1998, ApJL, 503, L131+

\bibitem[{{Stiele} {et~al.}(2007){Stiele}, {Pietsch}, {Haberl}, \& {for the
  XMM-Newton M 31 large program collaboration}}]{stiele07}
{Stiele}, H., {Pietsch}, W., {Haberl}, F., \& {for the XMM-Newton M 31 large
  program collaboration}. 2007, ArXiv e-prints, 711

\bibitem[{{Stiele} {et~al.}(2008){Stiele}, {Pietsch}, {Haberl}, \&
  {Freyberg}}]{stiele08}
{Stiele}, H., {Pietsch}, W., {Haberl}, F., \& {Freyberg}, M. 2008, \aap, 480,
  599

\bibitem[{{Str{\" u}der} {et~al.}(2001){Str{\" u}der}, {Briel}, {Dennerl},
  {Hartmann}, {Kendziorra}, {Meidinger}, {Pfeffermann}, {Reppin}, {Aschenbach},
  {Bornemann}, {Br{\" a}uninger}, {Burkert}, {Elender}, {Freyberg}, {Haberl},
  {Hartner}, {Heuschmann}, {Hippmann}, {Kastelic}, {Kemmer}, {Kettenring},
  {Kink}, {Krause}, {M{\" u}ller}, {Oppitz}, {Pietsch}, {Popp}, {Predehl},
  {Read}, {Stephan}, {St{\" o}tter}, {Tr{\" u}mper}, {Holl}, {Kemmer},
  {Soltau}, {St{\" o}tter}, {Weber}, {Weichert}, {von Zanthier},
  {Carathanassis}, {Lutz}, {Richter}, {Solc}, {B{\" o}ttcher}, {Kuster},
  {Staubert}, {Abbey}, {Holland}, {Turner}, {Balasini}, {Bignami}, {La
  Palombara}, {Villa}, {Buttler}, {Gianini}, {Lain{\' e}}, {Lumb}, \&
  {Dhez}}]{stru01}
{Str{\" u}der}, L., {Briel}, U., {Dennerl}, K., {et~al.} 2001, A\&A, 365, L18

\bibitem[{{Supper} {et~al.}(2001){Supper}, {Hasinger}, {Lewin}, {Magnier}, {van
  Paradijs}, {Pietsch}, {Read}, \& {Tr{\" u}mper}}]{S01}
{Supper}, R., {Hasinger}, G., {Lewin}, W.~H.~G., {et~al.} 2001, A\&A, 373, 63

\bibitem[{{Supper} {et~al.}(1997){Supper}, {Hasinger}, {Pietsch}, {Truemper},
  {Jain}, {Magnier}, {Lewin}, \& {van Paradijs}}]{S97}
{Supper}, R., {Hasinger}, G., {Pietsch}, W., {et~al.} 1997, A\&A, 317, 328

\bibitem[{{Titarchuk}(1994)}]{tit94}
{Titarchuk}, L. 1994, \apj, 434, 570

\bibitem[{{Trinchieri} \& {Fabbiano}(1991)}]{tf91}
{Trinchieri}, G. \& {Fabbiano}, G. 1991, ApJ, 382, 82

\bibitem[{{Trudolyubov} {et~al.}(2005){Trudolyubov}, {Kotov}, {Priedhorsky},
  {Cordova}, \& {Mason}}]{trud05}
{Trudolyubov}, S., {Kotov}, O., {Priedhorsky}, W., {Cordova}, F., \& {Mason},
  K. 2005, \apj, 634, 314

\bibitem[{{Trudolyubov} \& {Priedhorsky}(2004)}]{tp04}
{Trudolyubov}, S. \& {Priedhorsky}, W. 2004, \apj, 616, 821

\bibitem[{{Turner} {et~al.}(2001){Turner}, {Abbey}, {Arnaud}, {Balasini},
  {Barbera}, {Belsole}, {Bennie}, {Bernard}, {Bignami}, {Boer}, {Briel},
  {Butler}, {Cara}, {Chabaud}, {Cole}, {Collura}, {Conte}, {Cros}, {Denby},
  {Dhez}, {Di Coco}, {Dowson}, {Ferrando}, {Ghizzardi}, {Gianotti}, {Goodall},
  {Gretton}, {Griffiths}, {Hainaut}, {Hochedez}, {Holland}, {Jourdain},
  {Kendziorra}, {Lagostina}, {Laine}, {La Palombara}, {Lortholary}, {Lumb},
  {Marty}, {Molendi}, {Pigot}, {Poindron}, {Pounds}, {Reeves}, {Reppin},
  {Rothenflug}, {Salvetat}, {Sauvageot}, {Schmitt}, {Sembay}, {Short},
  {Spragg}, {Stephen}, {Str{\" u}der}, {Tiengo}, {Trifoglio}, {Tr{\" u}mper},
  {Vercellone}, {Vigroux}, {Villa}, {Ward}, {Whitehead}, \& {Zonca}}]{turn01}
{Turner}, M.~J.~L., {Abbey}, A., {Arnaud}, M., {et~al.} 2001, A\&A, 365, L27

\bibitem[{{van der Klis}(1994)}]{vdk94}
{van der Klis}, M. 1994, ApJs, 92, 511

\bibitem[{{van der Klis}(1995)}]{vdk95}
---. 1995, {X-ray Binaries} ({Cambridge University Press}), 256--307

\bibitem[{{Verbunt} {et~al.}(1995){Verbunt}, {Bunk}, {Hasinger}, \&
  {Johnston}}]{verb95}
{Verbunt}, F., {Bunk}, W., {Hasinger}, G., \& {Johnston}, H.~M. 1995, \aap,
  300, 732

\bibitem[{{Verbunt} \& {Lewin}(2006)}]{vl06}
{Verbunt}, F. \& {Lewin}, W.~H.~G. 2006, {Globular cluster X-ray sources}
  (Compact stellar X-ray sources), 341--379

\bibitem[{{White} \& {Angelini}(2001)}]{wa01}
{White}, N.~E. \& {Angelini}, L. 2001, ApJL, 561, L101

\bibitem[{{White} {et~al.}(1988){White}, {Stella}, \& {Parmar}}]{wsp88}
{White}, N.~E., {Stella}, L., \& {Parmar}, A.~N. 1988, ApJ, 324, 363

\bibitem[{{Williams} {et~al.}(2004){Williams}, {Garcia}, {Kong}, {Primini},
  {King}, {Di Stefano}, \& {Murray}}]{will04}
{Williams}, B.~F., {Garcia}, M.~R., {Kong}, A.~K.~H., {et~al.} 2004, ApJ, 609,
  735

\bibitem[{{Wilms} {et~al.}(2000){Wilms}, {Allen}, \& {McCray}}]{wilm00}
{Wilms}, J., {Allen}, A., \& {McCray}, R. 2000, \apj, 542, 914

\end{thebibliography}
\end{document}